\documentclass[aps,prl,twocolumn,superscriptaddress]{revtex4}
\usepackage{epsfig}
\usepackage{amssymb}
\usepackage{amsmath}
\usepackage{amsfonts}
\usepackage{graphicx}
\usepackage{color}

\begin{document}

\title{Non-Markovian polariton dynamics in organic strong coupling}
\date{ms.tex, \today}
\author{A. Canaguier-Durand}
\affiliation{ISIS \& icFRC, Universit\'{e} de Strasbourg and CNRS
(UMR 7006), 8 all\'{e}e Gaspard Monge, F-67000 Strasbourg, France.}
\author{C. Genet}
\affiliation{ISIS \& icFRC, Universit\'{e} de Strasbourg and CNRS
(UMR 7006), 8 all\'{e}e Gaspard Monge, F-67000 Strasbourg, France.}
\author{A. Lambrecht}
\affiliation{Laboratoire Kastler Brossel, ENS, UPMC and CNRS (UMR 8552), Campus Jussieu,
F-75252 Paris, France.}
\author{T. W. Ebbesen}
\affiliation{ISIS \& icFRC, Universit\'{e} de Strasbourg and CNRS (UMR 7006), 8 all\'{e}e
Gaspard Monge, F-67000 Strasbourg, France.}
\author{S. Reynaud}
\affiliation{Laboratoire Kastler Brossel, ENS, UPMC and CNRS (UMR 8552), Campus Jussieu,
F-75252 Paris, France.}

\begin{abstract}
Strongly coupled organic systems are characterized by unusually
large Rabi splittings, even in the vacuum state. They show the
counter-intuitive feature of a lifetime of the lower polariton state
longer than for all other excited states. Here we build up a new
theoretical framework to understand the dynamics of such coupled
system. In particular, we show that the non-Markovian character of
the relaxation of the dressed organic system explains the long
lifetime of the lower polariton state.
\end{abstract}

\maketitle

\section{Introduction}

Over the past 15 years, light-matter strong coupling has been studied
extensively with organic materials \cite%
{LidzeyNature1998,FujitaPRB1998,SchouwinkCPL2001,TakadaAPL2003,HolmesPRL2004,%
TischlerPRL2005,MichettiPRB2005,HakalaPRL2009,BellessaPRL2004,DintingerPRB2005,%
GuebrouPRL2012,AmbjornssonPRB2006,BerrierACSNano2011} which can
display very large splitting of the two hybrid light-matter states,
also known as the polariton states. Recently, optical resonances
with small mode volumes such as Fabry-Perot nanocavities or surface
plasmons have been used to achieve the so-called ultra-strong
coupling where the Rabi splitting approaching $\sim 1\mathrm{eV}$
becomes a significant fraction of the electronic transition energy
\cite{CuitiPRB2005,SchwartzPRL2011}. For such large splittings,
changes in bulk properties are observed, as already shown for the
work-function \cite{HutchisonAdvMat2013} and the ground state energy
\cite{CanaguierACIE2013}. It has also been noticed over the years
that the lifetime of the lowest polariton state, denoted
$\mathrm{C}^{-}$, is much longer than the lifetime of the photon in
the cavity mode
\cite{SongPRB2004,WiederrechtPRL2007,SalomonACIE2009,VasaACSNano2010,%
VirgiliPRB2011,HaoACIE2011,AgranovitchPRB2003,LitinskayaJLumin2004}.
In recent experiments using resonant excitation, this
$\mathrm{C}^{-}$ lifetime has even been shown to be longer than that
of the bare excited molecules
\cite{HutchisonACIE2012,SchwartzCPC2013}.

These properties are counter-intuitive in the conventional picture
where the dynamic properties of the coupled states are directly
determined from those of the bare ones \cite{WeisbuchJLumin2000}. In
the so-called Markov approximation, the effects of coupling and
relaxation are simply added to each other in the master equation
which describes the evolution of the system. It follows that the
relaxation rates in the diagram of dressed states are obtained from
those of bare states through a mere change of basis
\cite{CohenJPhys1977}. In the ultra-strong coupling limit in
particular, the low- and high-energy dressed states $\mathrm{C}^{-}$
and $\mathrm{C}^{+}$ contain identical proportions of the bare
states and their lifetimes are thus expected to be equal to each
other. The experimental observation of very different lifetimes for
these two dressed states reveals that the relaxation of the dressed
system is deeply influenced by the strong coupling. In other words,
the relaxation of coupled organic molecules corresponds to a
non-Markovian regime where relaxation can only be studied after the
effect of ultra-strong coupling has been taken into account
\cite{ReynaudJPhys1982}.

In the present article, we build up a new theoretical framework to
understand the dynamics of ultra-strongly coupled organic molecules.
In particular, we show that the relaxation inherent is intrinsically
non-Markovian in such a system. This new view on strongly coupled
organic materials explains the most salient features experimentally
observed in such systems, in particular the very long lifetime of
the lower dressed state $\mathrm{C}^{-}$.

\begin{figure}[tbp]
\centerline{\psfig{figure=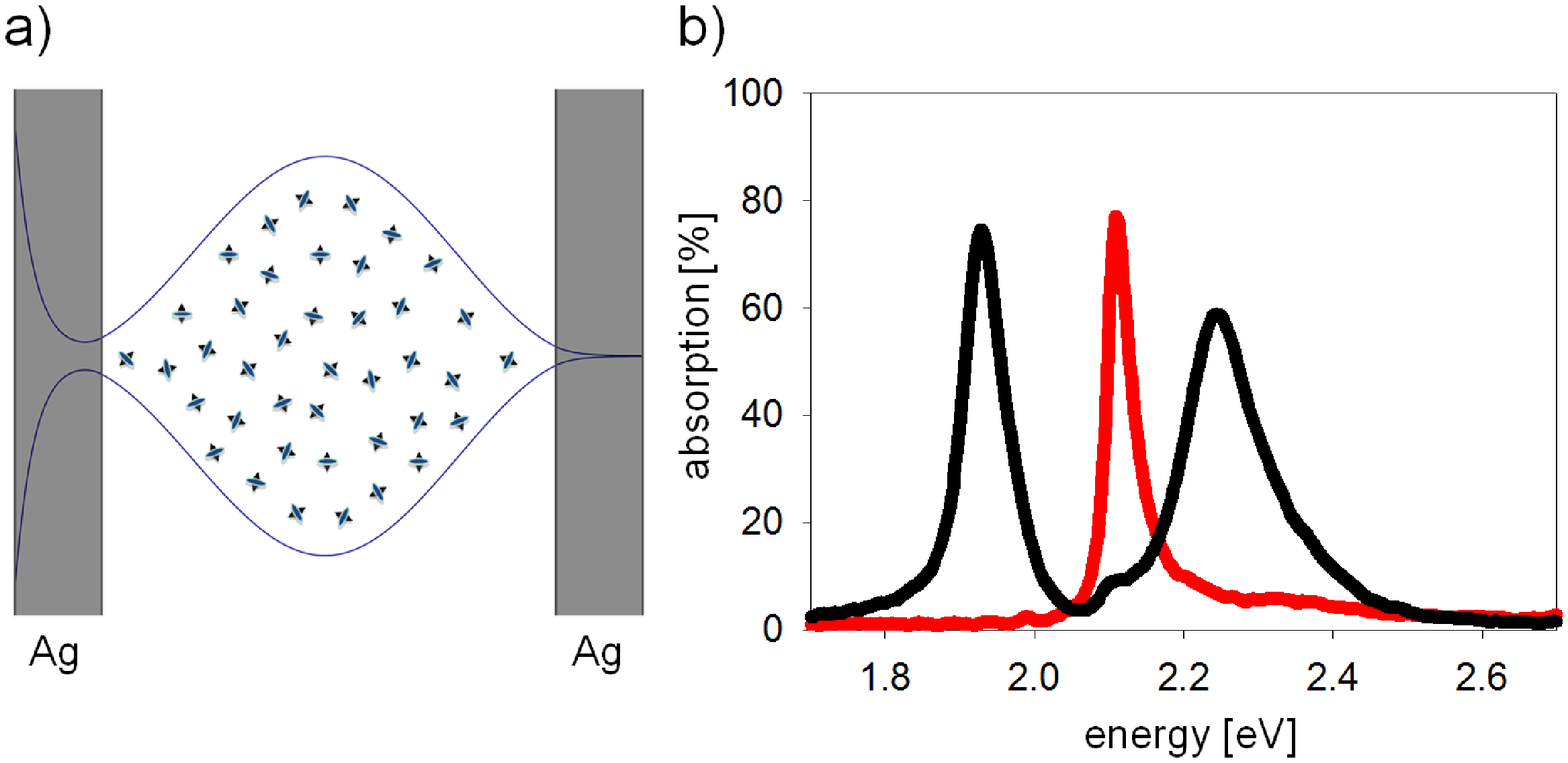,width=9cm}} \caption{a)
Illustration of molecules coupled to the fundamental optical mode of
a $145$nm thick Fabry-Perot cavity made of two $30$nm thick Ag
mirrors. b) Typical example of absorption spectrum of uncoupled (red
line) and coupled (dark line) molecules. The data correspond to
J-aggregate (TDBC) molecules dispersed in a polyvinyl alcohol (PVA)
polymer host matrix inside the cavity sketched in a)
\cite{SchwartzCPC2013}.} \label{fig1}
\end{figure}

\section{Non-Markovian dynamics}

Figure 1 illustrates the case of organic molecules strongly coupled
to a Fabry-Perot cavity mode. The organic molecules are doping a
host polymer matrix {at $0.1$ to $0.01$ molar concentration (mole
per litre). Qualitatively, this corresponds to typical
intermolecular separation distances of the order or larger than
$3$nm within the host matrix, so that F\"orster-type energy transfer
is expected to dominate over other intermolecular transfer
mechanisms \cite{ScholesARPC2003}}. Absorption spectra on the
right-hand side of the figure shows the effect of strong coupling
which splits the molecular resonance in the coupled system (dark
curve), as compared to the uncoupled one (red curve).

We stress here that the widths of the molecular absorption peaks
need not to be directly related to the intrinsic molecular
lifetimes, due to inhomogeneous broadening and vibrational manifold.
Inhomogeneous broadening is crucial for coupled and uncoupled
organic molecules to coexist in the cavity in the model discussed
further down, and this prevents one to draw conclusions about
intrinsic lifetimes from the measured spectral features.
Inhomogeneous broadening is due to distribution of orientations,
locations and micro-environment of the organic molecules in the
matrix. These features are essentially the same for coupled and
uncoupled molecules, since the optical coupling does not affect the
associated motions. In addition, the host matrix behaves as a
vibrational relaxation reservoir in thermodynamic equilibrium with
both coupled and uncoupled molecules.

The vibrational reservoir spectra are characterized by a typical
energy dispersion $k_{\mathrm{B}}T\simeq 25\mathrm{meV}$ at room
temperature, or equivalently a correlation time $\tau _{c}\simeq
\hbar /k_{\mathrm{B}}T$ $\simeq 25\mathrm{fs}$. The condition of
validity of the Markov approximation \cite{ReynaudJPhys1982} would
be that the Rabi splitting $\Omega _{\mathrm{R}}$ be inefficient
during the correlation time $\Omega _{\mathrm{R}}\tau _{c}\ll 1$,
that is equivalently $\hbar \Omega_{\mathrm{R}} \ll
k_{\mathrm{B}}T$. This condition is clearly not met for ultra-strong
coupling of organic molecules, where the Rabi splitting $\hbar
\Omega_{\mathrm{R}}$ is much larger than $k_{\mathrm{B}}T$
\cite{HoudrePRA1996,LienauNatPhot2013}. This implies that the system
is intrinsically in the non-Markovian regime, with relaxation
strongly influenced by the coupling. In other words, there is no
reason to expect that the dressed states $\mathrm{C}^{+} $ and
$\mathrm{C}^{-}$ have identical lifetimes, as it would be the case
for ultra-strong coupling in the Markovian approximation.
Furthermore, we will see below that the hierarchy of lifetimes
observed in experiments is naturally explained by the approach
proposed in this paper.

In our case, each individual molecule can only be weakly coupled to
the electromagnetic mode of the cavity. The strong coupling
mechanism necessarily involves a collective excitation of an
extremely large number of molecules coherently coupled to the single
mode of the cavity. It is important to stress that the strong
coupling does not shield the molecules from intramolecular
vibrational relaxation. This explains the extremely low emission
quantum yields, as observed experimentally, with vibrational
relaxation rates at least $100$ times larger than the radiative rate
of $\mathrm{C}^{-}$. For instance in the case of TDBC presented in
Figure 1, the fluorescence quantum yield of $\mathrm{C}^{-}$,
angularly integrated, is found to be $\sim4\times 10^{-3}$ (more
numbers will be given below).

We build up below the new framework which naturally allows us to
analyze such situations. We show in particular that the
non-Markovian character explains the otherwise counter-intuitive
long lifetime of the lower dressed state $\mathrm{C}^{-}$. We stress
at this point that a problem dominated by radiative relaxation would
lead to different conclusions \cite{CohenJPhys1977}. Note also that,
in what follows, dark states which are formed in the coherent state
manifold when coupling a large number of molecules to one cavity
mode are ignored \cite{HoudrePRA1996}. Nevertheless, their mere
presence in the energy diagram also contributes to the non-radiative
decay discussed further down.

\section{Bare and dressed states}

We consider uncoupled (U) and coupled (C) states as two populations
in a dynamical equilibrium with the total concentration
$[\mathrm{M}]=[\mathrm{U}]+[\mathrm{C}]$ fixed. This two-population
model is a simplification of the real situation where there is a
continuous distribution of molecules in the cavity mode with
different positions, orientations or environments which lead to
spectral inhomogeneous broadening. This model is based on recent
experimental observations which have been shown to be explained in
terms of two populations coexisting at thermal equilibrium with well
defined Gibbs free energies \cite{CanaguierACIE2013}. As usually,
the model has to be tested by comparing its predictions to
experimental observations.

The relevant states of the uncoupled molecules are, on the one hand,
the ground and excited states of the molecule $\mathrm{U}$ and
$\mathrm{U}^\ast$ and, on the other hand, the $0-$ and $1-$photon
states of the cavity. The states of the hybrid `` molecule+cavity''
system are denoted $\mathrm{U}_{0}$ for the ground state,
$\mathrm{U}_{1}$ and $\mathrm{U}_{0}^{\ast }$ for the excited ones
(see Figure 2). The energy difference between the two excited states
is
\begin{equation}
\hbar \delta = \hbar\omega _{1}-\hbar\omega _{\ast }~,
\end{equation}%
with $\omega _{1}$ the frequency of photons in the cavity mode and
$\omega _{\ast }$ the frequency of the molecular transition. The
detuning $\delta $ (as the Rabi coupling discussed in the next
paragraph) has a single value in the simplified two-population model
whereas it would have a distribution of values in a microscopic
description.

The relevant states are similar for coupled and uncoupled molecules,
with differences caused by the effects of the coupling. They are
denoted $\mathrm{C}_{0}$, $\mathrm{C}_{1}$ and $\mathrm{C}_{0}^{\ast
}$, with the symbol C replacing U. The excited states
$\mathrm{C}_{1}$ and $\mathrm{C}_{0}^{\ast }$ are coupled through
the Rabi coupling $2\upsilon$ which is not zero for the coupled
molecules. Note that this Rabi splitting has a large value, though
the cavity has a low quality factor $Q$ and remains in low states
with only 0 or 1 photon. This unusual feature is due to the already
discussed fact that the cavity field is coupled to a giant dipole
corresponding to the coherent superposition of an extremely large
number of molecules.

The dressed states, denoted $\mathrm{C}^{+}$ and $\mathrm{C}^{-}$,
are obtained by diagonalizing the effect of the Rabi coupling
between the states $\mathrm{C}_{1}$ and $\mathrm{C}_{0}^{\ast }$
\begin{eqnarray}
\mathrm{C}^{+}=\cos \theta ~\mathrm{C}_{0}^{\ast }+\sin \theta
~\mathrm{C}_{1} &&~, \nonumber \\ \mathrm{C}^{-}=\cos \theta
~\mathrm{C}_{1}-\sin \theta ~\mathrm{C}_{0}^{\ast } &&~,
\label{dressed}
\end{eqnarray}%
with the angle $\theta $ defined by
\begin{equation}
\tan (2\theta)=-2\frac\upsilon\delta \;,\quad 0\leq 2\theta \leq
\pi~.
\end{equation}%
$\mathrm{C}^{+}$ is defined to have a higher energy than
$\mathrm{C}^{-}$ and the splitting between the two states is
\begin{equation}
\Omega _{\mathrm{R}}=\sqrt{\delta ^{2}+4\upsilon ^{2}}~.
\end{equation}%
The projection factors in eq.(\ref{dressed}) are
\begin{equation}
\cos ^{2}\theta = \frac{\Omega _{\mathrm{R}}-\delta}
{2\Omega_\mathrm{R}} \;,\quad \sin ^{2}\theta =
\frac{\Omega_\mathrm{R}+\delta}{2\Omega _\mathrm{R}}~.
\end{equation}%
When the coupling is much larger than the detuning ($2\upsilon \gg
\left\vert \delta \right\vert $), these projection factors are
nearly equal $\cos ^{2}\theta \simeq\sin ^{2}\theta \simeq 1/2$.

\begin{figure}[tbp]
\centerline{\psfig{figure=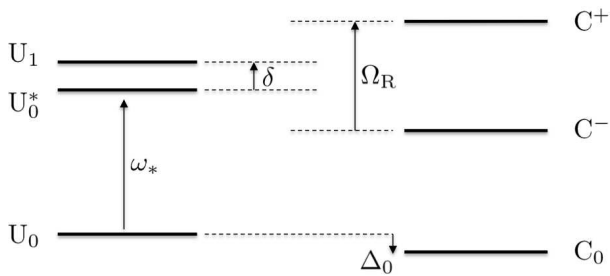,width=8cm}} \caption{Energy
diagram of the bare states $\mathrm{U}$ of the ``molecule$+$cavity''
system and of the dressed states $\mathrm{C}$. The energy difference
between $\mathrm{U}_{1}$ and $\mathrm{U}_{0}^{\ast }$ is $\hbar
\delta $, with $\delta =\omega _{1}-\omega _{\ast }$ the detuning
between the frequency $\omega _{1}$ of the cavity mode, and $\omega
_{\ast }$ that of the molecular transition. The Rabi splitting
$\Omega _{\mathrm{R}}$ between the dressed states $\mathrm{C}^{+}$
and $\mathrm{C}^{-}$ and the energy shift $\Delta_0$ of the ground
state $\mathrm{C}_{0}$ are shown.}
\end{figure}

All molecular states are connected in the {molecular} hamiltonian so
that the splitting of $\mathrm{C}^{+}$ and $\mathrm{C}^{-}$ has
consequences on the other states. This causes in particular a shift
$\Delta _{0}$ of the position of the ground state $\mathrm{C}_{0}$
\cite{FornDiazPRL2010,CuitiPRB2005,CanaguierACIE2013}. With an
observed Rabi splitting $2\upsilon\sim 1~\mathrm{eV}$ and a distance
between the ground and excited states $\Delta \sim 2~\mathrm{eV}$,
this shift cannot be neglected. A naive expectation $\left(
2\upsilon \right) ^{2}/2\Delta $ from second order perturbation
theory leads to a value consistent with the result of recent
measurements $\Delta _{0}\sim 0.1\mathrm{eV}$. Indeed, as this shift
changes the energy differences between the states of uncoupled and
coupled molecules, it can be measured in a thermodynamic approach as
the standard Gibbs free energy difference between the ground states
of the uncoupled and coupled molecules \cite{CanaguierACIE2013}.

\section{Cavity relaxation processes}

We now discuss the radiative relaxation processes which correspond
to emission of a photon by the cavity while leaving the molecular
state unaffected. The basis of the method is the application of
Fermi's golden rule to dressed states \cite{CohenJPhys1977}.

For the uncoupled molecules, there is only one relaxation channel
corresponding to the transition $\mathrm{U}_{1} \rightarrow
\mathrm{U}_{0}$. Simple rate equations describe the evolution of the
populations $[\mathrm{U}_{1}]$ and $[\mathrm{U}_{1}]$ due to this
process
\begin{equation}
\frac{\mathrm{d}[\mathrm{U}_{1}]}{\mathrm{d}t} = -
\frac{\mathrm{d}[\mathrm{U}_{0}]}{\mathrm{d}t} = -
\Gamma_{\mathrm{U}_{0}\mathrm{U}_{1}}[\mathrm{U}_{1}] ~,
\end{equation}%
and they preserve the sum of the two populations. The transition
rate $\Gamma _{\mathrm{U}_{0}\mathrm{U}_{1}}$, defined for the
transition $\mathrm{U}_{1}\rightarrow \mathrm{U}_{0}$, is the
product of a reduced rate $\gamma $ and a spectral density of
optical modes evaluated at the frequency of the transition.
Absorption rate on the same transition is the product of the
spontaneous emission rate $\Gamma _{\mathrm{U}_{0}\mathrm{U} _{1}}$
by a photon flux $\Phi _{\mathrm{U}_{0}\mathrm{U}_{1}}$ at the
relevant frequency
\begin{equation}
\frac{\mathrm{d}[\mathrm{U}_{1}]}{\mathrm{d}t} =
A_{\mathrm{U}_{1}\mathrm{U}_{0}}[ \mathrm{U}_{0}] \,,\;
A_{\mathrm{U}_{1}\mathrm{U}_{0}}=\Gamma _{\mathrm{U}
_{0}\mathrm{U}_{1}}\Phi _{\mathrm{U}_{0}\mathrm{U}_{1}}
\label{pumping}
\end{equation}
Note that the low $Q$ factor favors absorption events in the cavity
and thereby strong coupling.

For the coupled states, there are two radiative transition channels
$\mathrm{C}^\pm \rightarrow \mathrm{C}_{0}$ with rate equations
\begin{equation}
\frac{\mathrm{d}[\mathrm{C}^{\pm }]}{\mathrm{d}t} = -
\Gamma_{\mathrm{C}_{0}\mathrm{C}^{\pm }}[\mathrm{C}^{\pm }]~.
\end{equation}
The rates are proportional to squared projection factors $\Gamma
_{\mathrm{C}_{0}\mathrm{C}^{+}}\propto \sin ^{2}\theta$ and $\Gamma
_{\mathrm{C}_{0}\mathrm{C}^{-}}\propto \cos ^{2}\theta$, and to the
spectral densities of optical modes at the transition frequencies.
As these frequencies differ from the bare one, the values of the
emission and absorption rates differ from the expectations deduced
from the Markov approximation.

We note that the thermodynamical equilibrium is only slightly
modified by the absorption processes. The total population of
excited states does not exceeds a fraction of the order of $10^{-7}$
in the case of static spectroscopic experiments ($\sim 10^{-2}$ for
pump-probe measurements) so that the depletion of ground states
remains negligible. This means that the populations
$[\mathrm{U}_{0}]$ and $[\mathrm{C}_{0}]$ remain close to their
values in vacuum and also explains why stimulated emission processes
can be disregarded.

\section{Vibrational relaxation processes}

We now study vibrational relaxation processes which are the dominant
relaxation mechanism for most organic molecules. They correspond to
internal conversion of energy via a rapid cascade down the
vibrational ladder of the molecule. Typical organic molecules used
in strong-coupling experiments have over $100$ fundamental vibration
modes.

Another non-radiative relaxation process is the F\"orster energy
transfer between different molecules with conservation of energy.
Well known in molecular photophysics \cite{ScholesARPC2003}, these
processes correspond to a transfer of excitation due to F\"orster
dipole-dipole coupling between molecules over distances of a few nm
to a few tenths of nm. The energy excess, required for energy
conservation, is dissipated by a vibrational cascade down to the
lowest level of the corresponding electronic multiplicity, as
sketched on Figure 3. Though they involve Coulomb interaction, these
energy transfer mechanisms can be considered as non-radiative as
they do not couple to the free radiation field. It is also worth
noting that, at the small intermolecular distance scales where they
occur, they are expected not to perturb efficiently the coherence of
the collective dipole.

We do not enter into a detailed microscopic description of these
processes, well-known in molecular science, which leave the cavity
state unaffected. We give qualitative descriptions which are
sufficient for our purpose. A crucial feature in our case is that
the thermal energy $k_\mathrm{B}T$ is much smaller than energy
differences, so that downward transitions are dominant. The only
exception to this rule is the case of transitions between ground
states which correspond to a smaller energy shift $\Delta _{0}$ and
determine the thermodynamical equilibrium of the ground states of
the coupled and uncoupled molecules \cite{CanaguierACIE2013}.

For uncoupled states, there is only one non-radiative transition
$\mathrm{U}_{0}^{\ast }\rightarrow \mathrm{U}_{0}$. As previously,
this process is described by a rate equation
\begin{equation}
\frac{\mathrm{d}[\mathrm{U}_{0}^\ast]}{\mathrm{d}t} = -
W_{\mathrm{U}_{0}\mathrm{U}_{0}^\ast}[\mathrm{U}_{0}^\ast].
\end{equation}
The rate $W_{\mathrm{U}_{0}\mathrm{U}_{0}^\ast}$ is the product of a
reduced rate $w^\ast$ by a spectral density $\mathcal{S}$ which
represents the coupling of the two vibronic multiplicities and
depends on the energy difference. This reduced rate is relatively
small as this energy difference is much larger than $k_{B}T$. For
coupled states, there are similar transitions $\mathrm{C}^\pm
\rightarrow \mathrm{C}_{0}$
\begin{equation}
\frac{\mathrm{d}[\mathrm{C}^\pm]}{\mathrm{d}t} = W_{\mathrm{C}_{0}
\mathrm{C}^\pm}[\mathrm{C}^\pm] ~,
\end{equation}
with $W_{\mathrm{C}_{0}\mathrm{C}^{+}}$ and
$W_{\mathrm{C}_{0}\mathrm{C}^{-}}$ proportional to $\cos^2\theta$
and $\sin^2\theta$ respectively.

\begin{figure}[tbp]
\centerline{\psfig{figure=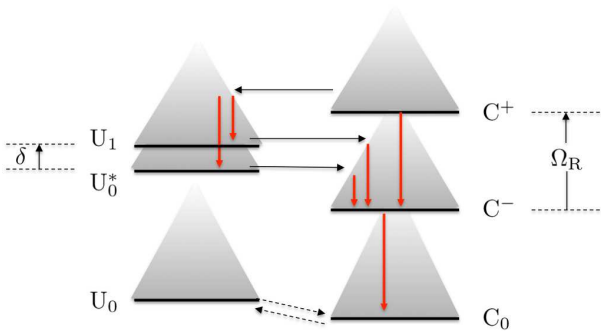,width=8cm}} \caption{Schematic
representation of uncoupled $\mathrm{U}$ and coupled $\mathrm{C}$
states in the non-Markovian regime. The vibrational ladders
associated with each molecular configuration are represented in grey
shadows and the non-radiative relaxation paths as red vertical
arrows. Transitions occuring between uncoupled and coupled molecules
are represented by horizontal black arrows. }
\end{figure}

There exists one relaxation channel which is opened by the strong
coupling and could never be seen in the absence of this effect. It
corresponds to the transition between the dressed excited states
$\mathrm{C}^{+}\rightarrow \mathrm{C}^{-}$
\begin{equation}
\frac{\mathrm{d}[\mathrm{C}^{+}]}{\mathrm{d}t} = -
W_{\mathrm{C}^{-}\mathrm{C}^{+}}[\mathrm{C}^{+}] ~,
\end{equation}
with a rate proportional to $\cos ^{2}\theta \sin ^{2}\theta $. This
new channel has a maximal rate when $\cos ^{2}\theta \simeq \sin
^{2}\theta \simeq 1/2$ and is very similar to the collisional
induced transitions studied in \cite{ReynaudJPhys1982}. Note that,
as the energy difference is smaller, the rate is larger than for
transitions studied in the preceding paragraph.

We come now to a second category of transitions occuring between
coupled and uncoupled molecules schematized in Figure 3. Such
transitions are observed experimentally as energy transfer processes
with well defined signatures \cite{VirgiliPRB2011,SchwartzCPC2013}.
In the study of ground states, we consider reverse transitions
$\mathrm{C}_{0}\rightarrow \mathrm{U}_{0} $ and
$\mathrm{U}_{0}\rightarrow \mathrm{C}_{0}$ because the energy
difference $\Delta _{0}$ is not so large with respect to
$k_{\mathrm{B}}T$. These transitions produce the thermodynamical
equilibrium between populations of coupled and uncoupled molecules
\begin{equation}
\frac{[\mathrm{C}_{0}]}{[\mathrm{U} _{0}]} =
\exp\frac{\hbar\Delta_0}{k_{\mathrm{B}}T} ~.
\end{equation}
This equilibrium favors coupled molecules for a downward shift
$\Delta _{0}>0$ of the coupled state.

For similar transitions between the excited states of coupled and
uncoupled molecules, energy differences are large, and we consider
only downward transitions $\mathrm{C}^{+}\rightarrow
\mathrm{U}_{1}$, $\mathrm{U}_{1}\rightarrow \mathrm{C}^{-}$,
$\mathrm{C}^{+}\rightarrow \mathrm{U}_{0}^{\ast }$, $\mathrm{U}
_{0}^{\ast }\rightarrow \mathrm{C}^{-}$. The rates
$W_{\mathrm{U}_{1}\mathrm{C} ^{+}}$,
$W_{\mathrm{C}^{-}\mathrm{U}_{1}}$, $W_{\mathrm{U}_{0}^{\ast }
\mathrm{C}^{+}}$, $W_{\mathrm{C}^{-}\mathrm{U}_{0}^{\ast }}$ are
respectively proportional to $\sin ^{2}\theta $, $\cos ^{2}\theta $,
$\cos ^{2}\theta $ and $\sin ^{2}\theta $ and to spectral densities
at the relevant frequencies. Hence, they can only be calculated on
the diagram of dressed states and are not determined by rates known
for bare molecules. This situation, typical for a non-Markovian
regime, is in sharp contrast with the Markov approximation where the
downward and upward rates would be similar.

\section{Orders of magnitudes}

Magnitudes of the various rates are known from the experiments (see
for instance \cite{SchwartzCPC2013}). The largest rate corresponds
to the radiative transition between uncoupled states
$\Gamma_{\mathrm{U}_{0} \mathrm{U}_{1}}\sim 4\times
10^{13}\mathrm{s}^{-1}$ which is strongly favored by the cavity. In
particular it is much larger than other radiative rates which have
values in the range of $10^{11}\mathrm{s}^{-1}$.

Large values are also obtained for non-radiative transition rates
between excited states
$W_{\mathrm{C}^{-}\mathrm{C}^{+}},W_{\mathrm{U}_{1}\mathrm{C}
^{+}},W_{\mathrm{U}_{0}^{\ast
}\mathrm{C}^{+}},W_{\mathrm{C}^{-}\mathrm{U}
_{1}},W_{\mathrm{C}^{-}\mathrm{U}_{0}^{\ast }}\sim
10^{13}\mathrm{s}^{-1}$, which arise as consequences of strong
coupling. The first one $W_{\mathrm{C} ^{-}\mathrm{C}^{+}}$ has a
dependence $\propto \cos ^{2}\theta \sin ^{2}\theta $ which makes it
large for molecules with a Rabi coupling larger than the detuning. A
similar discussion applies to the products of rates on the cascades
$\mathrm{C}^{+}\rightarrow \mathrm{U}_{1}\rightarrow \mathrm{C}^{-}$
and $\mathrm{C}^{+}\rightarrow \mathrm{U}_{0}^{\ast }\rightarrow
\mathrm{C}^{-}$. They correspond to two-step relaxation processes
$\mathrm{C}^{+}\rightarrow \mathrm{C}^{-}$ which are large when
$\cos ^{2}\theta \sin ^{2}\theta $ has its maximum value. These
processes offer possibilities to explain a selection of strongly
coupled molecules among a diverse population. The other
non-radiative rates have smaller values $W_{\mathrm{U}_{0}\mathrm{U}
_{0}^{\ast }}\sim 10^{12}\mathrm{s}^{-1}$ and
$W_{\mathrm{C}_{0}\mathrm{C} ^{-}}\sim 10^{12}\mathrm{s}^{-1}\gg
\Gamma _{\mathrm{C}_{0}\mathrm{ C}^{-}}$.

These orders of magnitude allows one to write down a simplified
system of rate equations. The largest absorption rate is indeed the
one $A _{\mathrm{U}_{1}\mathrm{U}_{0}}$ associated to the absorption
from $\mathrm{U}_{0}$ to $\mathrm{U}_{1}$ and the main relaxation
channel is then through non-radiative relaxation from $
\mathrm{U}_{1}$ to $\mathrm{C}^{-}$. The populations of the states
$\mathrm{C }^{+}$ and $\mathrm{U}_{0}^{\ast }$ remain negligible at
all times and can be ignored in the following simplified system of
solutions
\begin{eqnarray}
&&\left[ \mathrm{U}_{1}\right] \left( t\right) \simeq
\int_{0}^{t}\mathrm{d} t^{\prime }~e^{-R_{\mathrm{U}_{1}}t^{\prime
}} A _{\mathrm{U}_{1}\mathrm{U}_{0}} \left( t-t^{\prime }\right) [
\mathrm{U}_{0}] ~,  \\
&&\left[ \mathrm{C}^{-}\right] \left( t\right) \simeq
\int_{t_{0}}^{t} \mathrm{\ d}t^{\prime
}~e^{-R_{\mathrm{C}^{-}}t^{\prime }}W_{\mathrm{C}^{-}
\mathrm{U}_{1}}[\mathrm{U}_{1}]\left( t-t^{\prime }\right) ~, \notag
\end{eqnarray}%
where $R_{\mathrm{U}_1}$ and $R_{\mathrm{C}^-}$ are the total
relaxation rates for states $\mathrm{U}_1$ and $\mathrm{C}^-$
\begin{eqnarray}
&&R_{\mathrm{U}_{1}}\simeq \Gamma_{\mathrm{U}_{0}\mathrm{U}_{1}} +
W_{\mathrm{ C}^{-}\mathrm{U}_{1}} ~,  \\
&&R_{\mathrm{C}^{-}} \simeq \Gamma_{\mathrm{C}_{0} \mathrm{C}^{-}} +
W_{\mathrm{C}_{0}\mathrm{C}^{-}} ~. \notag
\end{eqnarray}%
The population of $\mathrm{U}_{1}$ follows the pumping rate
(\ref{pumping}), with a delay determined by $R_{ \mathrm{U}_{1}}$.
As already stated, the population of $\mathrm{U}_{0}$ is not
significantly depleted and can be considered as constant. The
population of $\mathrm{ C}^{-}$ follows the feeding from
$\mathrm{U}_{1}$, with a delay determined by $R_{\mathrm{C}^{-}}$.
As $R_{\mathrm{U}_{1}}\sim 5\times 10^{13} \mathrm{s}^{-1}$ is
$\sim50$ times larger than $R_{\mathrm{C}^{-}}\sim
10^{12}\mathrm{s}^{-1}$, it follows that $[\mathrm{U}_{1}]$ reaches
a quasi-stationary value $A _{\mathrm{U}_{1}\mathrm{U}_{0}}
[\mathrm{U}_{0}] / R_{\mathrm{U}_{1}}$ after a very short time
$R_{\mathrm{U}_{1}}^{-1}\sim 20\mathrm{fs}$. Then $[\mathrm{C}^{-}]$
shows a quasi-stationary behavior for a much longer time $
R_{\mathrm{C}^{-}}^{-1}\sim 1\mathrm{ps}$ during which it is by far
the most populated excited state and determines all observables.
This explains the main feature observed in the experiments, that is
the extremely long lifetime of the lower dressed state
$\mathrm{C}^{-}$, which is much longer than that of other excited
states.

\section{Discussion}

The decay of $\mathrm{C}^{-}$ is dominated by the internal
vibrational relaxation whereas the radiative decay (fluorescence) is
a negligible pathway. Even if the {fluorescence} rate is not
suppressed, it is {overwhelmed} by the non-radiative rate enhanced
in the strong coupling regime due to internal conversion via
vibrational overlap between $\mathrm{C}^{-}$ and $\mathrm{C}_{0}$.
This increase of the non-radiative decay with respect to the
radiative one is confirmed by the small emission quantum yield
measured at the level of strongly coupled molecules (numbers given
below).

Meanwhile, the higher dressed state $\mathrm{C}^{+}$ is much shorter
lived due to the extremely rapid vibrational decay to
$\mathrm{C}^{-}$ and energy transfer to uncoupled molecules (see
Fig.3). The lifetime of $\mathrm{C}^{+}$ turns out to be less than
$150$ fs while the lifetime of $\mathrm{C}^{-}$ is of the same
order, at resonance, than that of the bare molecule
\cite{SongPRB2004,HutchisonACIE2012,SchwartzCPC2013}. {In fact,} the
strong dissymmetry in the $\mathrm{C }^{-}$ and $\mathrm{C}^{+}$
lifetimes is a direct proof of the importance of the vibrational
coupling for the decay process of the polaritons, as well as of the
non-Markovian character of the associated relaxation. As also known
for the lowest excited level of most molecules, $\mathrm{C}^-$ has a
very long lifetime precisely because the vibrational overlap between
the lowest excited level and the ground state is much smaller than
between it and the higher excited states.

Let us discuss here two examples.  For merocyanine strongly coupled
($\Omega _\mathrm{R}\sim 0.7$eV) to a Fabry-Perot cavity of low
$Q-$factor ($\sim 10$), the half-life of $\mathrm{C}^{-}\sim 10$ps
is much longer than the photon lifetime in the bare cavity ($\Gamma
_{\mathrm{U}_{0} \mathrm{U}_{1}} ^{-1} \sim 25$fs) while being
shorter than that of bare molecules ($30$ps)
\cite{HutchisonACIE2012}. In the case of the TDBC J-aggregate
strongly coupled ($\Omega _\mathrm{R}\sim 0.35$eV) to a similar low
$Q$ cavity, $\mathrm{C}^{-}$ has a half-life of $4$ps, which is even
longer than the $1$ps half-life of the bare organic material, as
shown in Fig.4. Note that these lifetime values are the same whether
$\mathrm{C}^{-}$ is excited resonantly or not. When the pump reaches
higher electronic levels of uncoupled or coupled molecules, the same
transient spectrum and lifetime are observed, confirming that
$\mathrm{C}^{-}$ determines the observable because the population
accumulates in this longest-lived state.

We also emphasize that for both types of molecules, the quantum
yields in the strong coupling regime are remarkably low. Indeed, for
merocyanine, a highly efficient organic dye, the measured quantum
yield associated with $\mathrm{C}^{-}$ falls below $10^{-4}$
\cite{HutchisonACIE2012,SchwartzCPC2013}. For TDBC, we measure a
quantum yield $\sim 4\times 10^{-3}$ \cite{WangPrep}. In fact, this
can also be found by simply remembering that for molecules with high
oscillator strength such as merocyanine and TDBC, the radiative rate
is at best $\sim10^{9}$s$^{-1}$ so that, given the observed
$\mathrm{C}^{-}$ life times of the order of picoseconds, quantum
yields are expected to be less than $10^{-2}$.

\begin{figure}[tbp]
\centerline{\psfig{figure=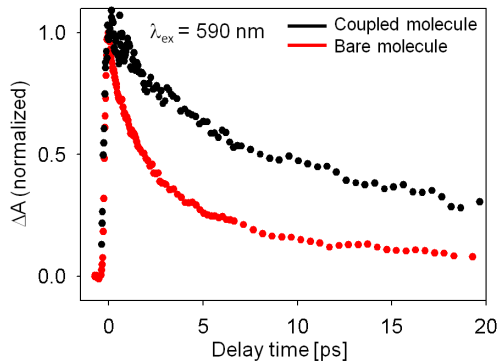,width=7cm}} \caption{Temporal
evolution of the total change in absorption recorded immediately
after a $150$fs pump pulse at $590$nm for a bare film of TDBC
molecules (red data) and for TDBC molecules coupled to the cavity
(black data; see Fig.1 b). After the pumping rise time, the
relaxation appears exponential over this time scale, in agreement
with Eq.(3). The half-live of $\mathrm{C}^{-}$ $\sim 4$ps (black) is
thus longer than that of the bare molecule (red). At this time
scale, the radiative lifetime ($\sim 25$fs) of the low $Q$ cavity
appears as instantaneous.} \label{fig4}
\end{figure}

\section{Conclusion}

In conclusion, we have shown in this paper that the observed lifetimes of
the polariton states are naturally explained in the non-Markovian
relaxation approach studied in the present letter. The lifetimes of
the excited states are determined by vibrational relaxation
phenomena and they are strongly affected by the large Rabi splitting
which changes the overlaps of the vibrational reservoirs. In
particular, the lifetime of the lower dressed state $\mathrm{C}^{-}$
is much longer than that of other excited states and its value is
disconnected from that of the photon decay rate in the bare cavity,
or of the relaxation rates of bare molecular states. This explains
the main features observed in experiments and also opens new
possibilities to influence chemical dynamics by controlling organic
strong coupling.

\paragraph{Acknowledgment -}

The authors are grateful to Claude Cohen-Tannoudji, James A.
Hutchison and Jean-Marie Lehn for fruitful discussions. This work
was funded in part by the ERC (grant 227577) and the ANR (Equipex
Union).

\end{document}